\documentclass[12pt]{article}
\usepackage[margin=2.5cm]{geometry}
\usepackage{graphicx}

\begin{document}
\title{Sensitivity Analysis of Ruin of an Insurance Company in Ghana}
\author{Daniel Tawiah Pabifio}
\date{November 30, 2022}
\maketitle
\noindent\rule{16cm}{0.4pt}

\begin{abstract}
An insurance company, as a risk bearer, is exposed to the likelihood of running into ruin. This is the situation where the initial surplus falls below zero. There is the need to find the required start-up capital to hedge against insolvency. Most researchers, irrespective of whether the test for claim dependency holds or not, assume claim independence in their computing of ruin probabilities to start up their initial capital. The objective of this study is to carry out comparative sensitivity analysis of ruin probability under both assumptions of dependence and independence, irrespective of whether the data exhibits independence or not, based on data from an insurance company in Ghana. Secondary data from an insurance company was obtained from the National Insurance Commission (NIC) for the period of 2013 to 2017. The study employed copulas to determine the claim dependence among the various insurance products and the company in general. Bowers \cite{bowers1997actuarial} used ruin probabilities to determine the ruin probabilities at various start-up capitals. The study concluded that when there is dependence in the claim data, computing the ruin probability based on the assumption of independence results in underestimation. Among the various insurance products, the most profitable insurance product was motor insurance, and Fire and Allied insurance exhibited the highest dependency. At a higher start-up capital, when claims are dependent, assuming independence in calculating the ruin probability results in a significant difference. Hence, it was recommended that insurance companies should adopt the assumption of dependence between the claims data as the initial reserves become larger, particularly for larger insurance companies, to avoid misleading results. Also, awareness of the perils and consequences of fire outbreaks and disasters should be raised to the general public to reduce the risk of frequent occurrence.
\end{abstract}

\section{INTRODUCTION}
	\subsection{Background Study}
	
Life is characterized by hazards of various types, some threatening all people and others limited to estate owners, while others are typical of some people or unique occupations. Insurance was born to take all or a portion of the risks from the risk-bearer. Under an insurance contract, the insurer agrees to pay a portion or all of the loss of the insured when there is the occurrence of an uncertain specified event or risk covered by the agreement. The policyholder agrees to pay an agreed sum of money, called a premium, on a regularly agreed basis into a pool in return for this protection or cover. Those who actually suffer the expected loss are compensated from that pool.
Insurance cover or security is achieved through a pooling system by which many people who are susceptible to a common danger are combined into a common risk pool. The insurer absorbs losses incurred over a specified period much easier by pooling the premium payments and the risk of a large number of policyholders. By doing so, they hedge risks better than an uninsured individual \cite{glied2017future}.
However, insurance companies run the risk that their surplus will become negative as they provide insurance cover. Therefore, insurers must take steps to decrease the likelihood that their surplus will become negative, that is, the probability of ruin. These measures include the investment of the surplus, taking reinsurance, portfolio selection, and volume control through the setting of premiums.
In analyzing the likelihood of the company making a profit or loss, there is the need to take into consideration the likelihood of ruin. This is more important to insurance companies because of the unpredictability of events that lead to claims and also the erratic nature of claim sizes. Ruin theory is therefore so important to the insurer, and it uses mathematical and statistical models to analyze the vulnerability to ruin of the insurer \cite{tang2019interplay}.
Insurance companies provide a source of protection to the people they insure. However, insurance covers only the consequences that can be measured in monetary terms. There may be other losses that may not be able to be measured in monetary terms and thus are not insured. Moreover, another limitation of insurance protection is the inability to directly decrease the loss probability. For example, the fact that there is a fire insurance policy does not necessarily alter the likelihood of a fire to occur. But there are well-designed systems of insurance that provide incentives for the protection of loss activities.

Every company is at the risk of failure and this is especially true for insurance companies whose primary purpose is to cover the risk of others. When the risk of failure is not well managed, the company can run into ruin. An insurance company has run into ruin if its reserve or capital is depleted to the point where it is unable to cover its insurance liabilities \cite{chen2015ruin}. Insolvencies are a major problem and there are reasons why they must be of major concern. Insolvencies of a company are usually not the result of one cause. More often than not, insolvencies occur due to a combination of factors.

\newpage

According to the National Insurance Commission (NIC), insurance products may be categorized as fire, burglary and damage to property, accidents, floods, marine and aviation, motor and general liability for non-life business and universal life, funeral, whole life, endowment, term, and group life for life business.

	\subsection{Problem Statement}
	
Insurance companies must have sufficient initial capital to make sure that the probability of ruin is minimal \cite{sattayatham2013ruin}. One reason why an insurance company may become insolvent is due to exposure to a catastrophe. This could be due to a greater number of claims from one event (an example is the June 3rd, 2015 flooding of Accra) or a modest number of big claims (the destruction of a very big building by fire can be an instance). Unforeseen claims can also threaten the insurance company and ultimately lead to ruin. The terrorist attacks on the World Trade Centre in 2001 were unforeseen by many, and many companies in the industry did not allow or charge for premiums on that. Huge losses were suffered by some companies and even led to the insolvency of Taisei Marine and Fire Insurance \cite{okura2016capital}. As a result of these uncertainties, there is the need for an insurance company to set aside enough reserves to cover the cost of all claims that arise in the business.
It is important to estimate the ruin probability because it safeguards the insurance company against different kinds of fluctuations. The fluctuations are the total claims on one hand and the total premiums on the other hand. However, the dependence among the claim sizes and claim counts is not considered when estimating the probability of ruin and setting the initial surplus to hedge against the probability of ruin \cite{kizinevivc2018exponential}.
According to Jaffee \cite{jaffee2015catastrophe}, the number of claims due to natural disasters and other catastrophes can contribute to the ruin of an insurance company. When the dependence between the claims has been tested and the ruin probabilities have been calculated, are they significantly different from the ruin probabilities under the assumption of independence between the claims? Statistical tests to test the significant difference between the two sets of ruin probabilities are usually not conducted to validate the assumptions.
Since statistical tests are not conducted to test the significant difference, the insurance company’s assigned initial surplus without the consideration of the right claims dependency of the company might lead to misleading results. There are several kinds of dependence to be considered: dependence within claims, between claim sizes, and inter-arrival times. The study will carry out comparative sensitivity analysis of ruin probability under both assumptions of dependence and independence, irrespective of whether the data exhibits independence or not, based on data from an insurance company in Ghana.

\newpage

	\subsection{Objectives of the Study}
	
The general objective of the study is to carry out a comparative sensitivity analysis of ruin probability under both assumptions of dependence and independence, irrespective of whether the data exhibits independence or not, based on data from an insurance company in Ghana. The specific objectives are:
\begin{itemize}
	\item To test for dependence or otherwise among the claims of different insurance products of the company.
	\item To fit appropriate models for the claims of the different products.
	\item To compute ruin probabilities for the different products under both types of dependency assumptions at different start-up capitals.
	\item To test for the existence of significant difference between the ruin probabilities under the two types of dependency assumptions.
	\item To identify which insurance product of the Ghanaian company performs better.
\end{itemize}

	\subsection{Organization of the Study}
	
The study starts with the introduction—touching the background study, problem statement and the objectives of the study. This is followed by the literature review on ruin theory, concept of surplus and premiums and claims as pertains to an insurance company. The next section reveals the data source and discusses the research methods used; which is then followed by analysis and discussions. The study concludes with conclusions and recommendations which may be considered for further study.

\newpage

\section{LITERATURE REVIEW}
	\subsection{Concept of Ruin Theory and Surplus}
		\subsubsection{Ruin Theory}
		
Ruin is the situation where the reserve falls below zero. This typically occurs after premiums have been received and claims have been paid. Ruin theory in actuarial science takes into consideration statistical methods and mathematical models to analyze the insurance company's propensity for running into ruin.

The vulnerability of an insurance company to ruin can be defined using ruin theory that utilizes mathematical models. A company is considered ruined if the surplus becomes negative. Ruin analysis is very necessary for the long-term planning and maintenance of the insurance company.

Risks faced by individuals and businesses are pooled together so that, in the event of a loss, the insurer (insurance company) shall compensate individuals and businesses in order to reduce the financial burden. When the specifically insured events happen, the insured will be entitled to claim a portion or all of the loss. An amount called a premium is paid to the insurance company in return for this claim. The insurance company is bound by law to honor its promise when they come due.

To guarantee that the insurance firm can pay out its claims as promised, it must set aside an amount that it can rely on whenever claims are due. This is called a reserve or surplus. Reserve, as the name suggests, is not gathered overnight but over long periods of time. This may possibly be from excess premiums received over claim amounts paid. In the modern business environment, surplus has many sources other than the traditional or primary method. One other source of surplus is investment income. Ignoring the impact of interest has been the traditional method.

Ruin probability is the probability that the insurance company’s surplus or reserve will become negative.

			\subsubsection*{Importance of Ruin Theory}
			
Ruin Theory is important for the following reasons:			
\begin{itemize}
	\item In the subsequent event of properly managing the realized risk of the insurance company, the company can be well protected from insolvency.
	\item Ruin Theory helps to quantify the apparent risk of collapse that insurance companies face.
\end{itemize}

		\subsubsection{Concept of Surplus}
	
Initial Surplus is the amount of money the insurance company sets aside as a reserve. This serves as a financial cushion in situations where the claims paid exceed the premiums received in the first year. The insurance surplus exists for the same purpose as in other companies: it acts as a financial buffer to protect against detrimental business circumstances in which operating losses occur. Surplus offers a cushion to cover losses, at least temporarily, and allows the company to continue operating as usual.

However, insurance is unique in that, at the moment the product is priced and purchased, a significant part of its business expenses (i.e., claim payments) cannot be determined. In essence, for many years, these expenses may not be known. Prolonging the uncertainty, many variables restrict the capacity to forecast these expenses with a large degree of certainty, such as social inflation and altering tort law. As a result, the correct amount of surplus necessary to support insurance writings is hard to determine. This section presents the concept of Bench surplus, Surplus Run-off and sources of surplus.

			\subsubsection*{Benchmark or Standard Surplus}
Benchmark surplus or standard surplus is that excess level that offers a business line or company segment a suitable financial buffer. The magnitude of the standard surplus for a business line must be based on an account of variables distinctive to that line that introduce uncertainty (or volatility) in anticipated future outcomes. It should also reflect the probability of occurring detrimental conditions that would cause an excess drain. The higher the amount of surplus, the less likely it will reduce the whole volume of a company's surplus when detrimental circumstances occur. The notion of the probability of adverse circumstances occurring is essential to a standard surplus being established. With a defined probability of financial ruin, a quantity of standard surplus is regarded hand in hand. Standard surplus is not mandatory surplus nor GAAP equity. Standard surplus is a system for financial cushioning that is essential and may or may not match the reported surplus of a specific company. However, it reflects the realities a business should consider in its operating procedures.

			\subsubsection*{Surplus Run-off}
Using a premium-to-surplus ratio, Surplus Run-off expresses the necessary excess in relation to premium. Using this ratio should not conceal the fact that while premium flows usually span a single year, there are demands for surplus for policy cash flows across the whole run-off period, no matter how long it may take. The benchmark excess requirement continues beyond the year the business is created. It is recommended that over time, surplus committed to supporting the company can decrease in relation to the decrease in funding. Just as funding is the present-valued assets corresponding to future cash flows, which decline over time, the necessary surplus should be regarded as the related present-valued assets running off in parallel fashion. Because loss reserves are mainly the primary element of this reserve financing requirement, this implies that surplus would decrease as loss reserves are reduced to zero in a simpler way. The premium-to-surplus ratio's comfort and simplicity promote its extensive use. A reserve-to-surplus ratio would be much more relevant than a premium-to-surplus leverage statistic and would provide a more intuitive means of allocating surplus. The technique shown here is to provide an estimate of standard initial surplus demands using average payment dates. Insurance programs with an atypical cash flow pattern may involve a more comprehensive cash flow model in order to estimate the overtime surplus requirements. A multi-risk insurance company may bundle together a greater leverage.

		\subsection{Sources of Surplus}
		
Surplus is not necessarily the outcome of accumulated earnings. Since surplus is merely the balancing factor that shows the difference between assets and liabilities plus the capital stock, any revaluation of the assets without a corresponding increase in liabilities or any reduction in liabilities without a corresponding reduction in the amount of the assets results in increased surplus. Although profits constitute the predominant source of surplus, there are many other sources of surplus. The management must have a clear comprehension of these sources so that they may accurately determine the real worth of the surplus \cite{dickson2016insurance}.

A detailed and careful analysis of different sources of surplus is necessary when the declaration of dividends is contemplated. Although directors may have the legal right to declare a dividend from any surplus, it is not always expedient or good financial policy to do so. All surplus should be administered in the company's best interest, and wide discretion may be used in handling it. The main types of surpluses are earned surplus, paid-in surplus, and unrealized appreciation of assets. There are a number of sources of these types of surpluses. Besides discussing the nature of these sources of surplus, we shall attempt to examine which source can be gainfully used for dividend purposes.

			\subsubsection{Sources of Earned Surplus}

Capital surplus arises from every possible source except earnings from normal operations of the firm. It may be paid-in surplus, donated surplus, appreciation of assets, mergers and consolidations, sale of assets in excess of book value, and reduction in stated capital.

			\subsubsection*{Sale of Par Stock}

When a stock is sold at more than its par value, the excess or premium is credited to paid-in surplus \cite{glied2017future}. This source of surplus should not ordinarily be used for dividends. The securities were sold to raise capital, and the entire proceeds of the sale should really belong to the capital. Sometimes securities are sold at a premium for the specific purpose of using for dividends the surplus created by premiums.

			\subsubsection*{Sale of No-par-stock}
			
The sale of no-par stock constitutes a potential source of paid-in surplus \cite{okura2016capital}. When a company sells such stock, it transfers a portion of the sale proceeds to the capital stock account and the remainder to paid-in surplus. Such funds are part of the capital of the company, and hence dividends should not be distributed from this source.

			\subsubsection*{Conversion of Debentures}

When a bond or a preferred stock is converted into a common stock having a less par or stated value, paid-in surplus is created \cite{glied2017future}. Since this activity does not release any fund, this surplus cannot be used for dividend.

			\subsubsection*{Reduction in Par Value}
Surplus is sometimes created through a reduction in the par value of a company’s stock and a reduction in the stated value of the company's non-par stock \cite{glied2017future}. Surplus derived through such recapitalization processes should not be used for dividend purposes because this will impair the capital of the company.

			\subsubsection*{Forfeited Subscriptions and Cancellation Indebtedness}
			
When a subscriber for stock fails to meet their obligations, the company forfeits the payments already made by the subscriber. A reduction in liabilities due to cancellation of indebtedness or a compromise with creditors would add to book surplus. As the surplus does not release any cash to the company, the management should not use it for dividend distribution \cite{afonso2013dividend}.

			\subsubsection*{Surplus from Purchase of Capital Liabilities}
			
When a company buys its own bonds and stocks below par or the value at which the stock is carried on the liability side of the balance sheet, paid-in surplus is created. Funds obtained from the original sale of such securities are part of the capital of the company, and hence dividends should not be paid from a surplus arising in this manner \cite{afonso2013dividend}.

\newpage

		\subsection{Premiums and Claims}
		
Premium is a consideration amount paid periodically by the policyholder to the insurance company to cover the risk transferred to the insurer. The value of the premium is ascertained after a financial decision rule has been accepted by both parties.

Life insurance and non-life insurance have different forms of premiums. Life insurance premiums or payments are characterized by long-term and periodic (monthly, quarterly, or annually) payments. By making an accepted advance or set amount of cash, the insured initiates the policy. When there is a well-defined random event, specified payments made by the policyholder to the insurance company with an agreement to provide financial coverage against a loss are known as the premium.

The policyholder’s right to the sum stipulated by the agreement is referred to as a claim or loss amount. The amount of the claim is the amount to be paid by the insurer in the case of an occurrence. There are generally time lapses between the occurrence of a case and the settlement of claims. Also, the time it generally takes to report events is a factor. For example, a claim for property insurance may be resolved more quickly than a claim for a motor accident because it involves a certain amount of time to determine the degree of injury or harm.

Upon receiving premiums, the insurer, as a business, puts aside a surplus to meet future claims. The insured is entitled to a claim on the insurance company if the event, which contracted the insurance policy, occurs. Claims are claims for payment under the terms and conditions of an insurance contract by a policyholder or an alleged third party; technically, the claimant is the policyholder or third party who requests payment. It is the insurer's responsibility to ensure that, once they are applied for, adequate funds or reserves are available to pay claims. This is mainly the reasoning behind learning the average amount to be paid out in a given year. Depending on the premiums taken, the reserved sums should be enough.

The surplus of the claims is referred to as claims reserve; these are the exceptional claims for the compensation of future policy owners that have already occurred. These essential amounts are laid aside by actuaries for claims not yet disclosed to the company (referred to as Incurred but Not Yet Reported (IBNYR)). Other sums constitute allowances for adjustments in the estimate of claim handlers and are known as Incurred but Not Enough Reported (IBNER). According to Mikosch (2009) and Mario and Wuthrich \cite{wuthrich2008stochastic}, most businesses merge these two amounts as Incurred but Not Reported (IBNR). How the company's liabilities are implemented remains fully random in nature compared to the assets (premiums and volumes of policies sold); therefore, care must be taken to consider the dynamics of the liabilities. When settling claims based on premiums and policies sold, the difference between the property processed and the rendered liabilities is known as excess processing. Managing these surpluses properly determines an insurance company's success or downturn. 

		\subsection{Solvency in Ghana}
		
The insurance industry's main objective is to ensure that insurance companies function on a financially sound basis that will contribute to effective resource allocation, effective risk management, promotion of economic growth, and long-term savings mobilization. There is a need for sound macroeconomic policies that are crucial for the efficient results of the insurance supervisory system for the insurance industry to profit and better safeguard policyholders. In 2008, the National Insurance Commission (NIC) pushed for a solvency system to deny insurance firms with poor financial standing the chance to conduct larger company operations as a measure to protect policyholders and ensure safe transaction handling \cite{nic2008national}.

		\subsection{Previous Literature on Ruin Theory}

Ruin theory started with Filip Lundberg in 1903. This collective risk model has become one of the most important tools for insurance actuaries in the world. Several works on the surplus of insurance companies have been performed in continuous time in risk theory. Such a model is more natural when describing the real world. However, the discrete process of financial surplus is considerably more modest.

In recent times, not only do the premium collection and the claim payment influence the surplus. Investment and reinsurance also play a significant role. The analysis of the investment of financial surplus enhances the security of an insurance company. Gatto and Baumgartner \cite{gatto2016saddlepoint} studied four different methods to calculate the insurance company's likelihood of ruin. The first approach was based on the saddle-point approximation of asymptotic analysis. Recursive upper and lower approximations was the second approach. The third approach was the fast Fourier transform, and the fourth approach was the Monte Carlo importance sampling. To illustrate the high performance of the four techniques, they performed numerical studies.
In the more extreme case, evaluating risk can be undertaken in a more practically realistic manner \cite{constantinescu2013ruin}. Risk management problems could bring contemplation from the classical model. These simple modifications could be made to make way for other considerations such as dividends, dependent stochastic interest rates, and even non-constant premiums to influence the dynamics.
In the classical ruin probability model, interest rates are assumed to be constant and hence independent. However, assuming that the interest rate is independent and identically distributed, as Cai \cite{cai2002expected} studied, is not realistic because we know that interest rates are statistically dependent over time. The dependent interest rates were modeled in an autoregressive structure. Using two generalized discrete-time risk processes, they calculated the ruin probabilities. Recurring and integral equations were provided for the ruin probability. Inductive and martingale approaches were used to derive the Lundberg inequalities for the ruin probabilities:

\begin{equation}
U_k=u\prod\limits_{j=1}^{k} (1+I_j)+ \sum\limits_{j=1}^{k} (X_i-Y_i) \prod\limits_{t=j+1}^{k} (1+I_t).........k=1,2,...
\end{equation}

Considering $\{X_n, n=1,2,...\}$ and $\{Y_n, n=1,2,...\}$ to be two sequences of independent and identically distributed non-negative random variable.

\begin{equation}
\Psi(u)= Pr\{\bigcup\limits_{k=1}^{\infty}(U_k < 0)\}
\end{equation}

where,
\begin{equation}
U_k = u + \sum\limits_{t=1}^{k} (X_t - Y_t),  k=1,2,...
\end{equation}
Or the stochastic process $\{U_n = 1,2,..\}$ satisfies \\
\begin{equation}
U_n = U_{n-1} + X_n - Y_n ,   n=1,2,...
\end{equation}

with $U_0 = u\geq 0 $ \\
The quantity $\Psi(u)$ has appeared in many applied probability models and has been studied extensively in ruin theory.$\Psi(u)$ has been regarded as the ruin probability in different risk models.
Ever since the inception of insurance, just like other financial institutions, insurance has undergone modifications, and one of them is reinsurance. Reinsurance is when an insurance company purchases insurance from another insurance company, transferring part of the insured liabilities to the reinsuring company. Reinsurance has a significant influence on increasing the security of an insurance company. In view of this, proportional reinsurance can be used to control the risk model\cite{diasparra2009bounds}. It has become a very active and protective area for study to control a risk process. However, there is a tough hurdle. Thus, an alternative method is commonly used—deriving inequalities or bounds for the ruin probabilities. The objective is to select methods for reinsurance control to bound the probabilities of ruin.

Aside from the insurance company’s reserve boosting the security of the company, it can receive interest upon investment \cite{kluppelberg2001developments}. Upon considering large claims, the results obtained applied to Pareto and log-gamma stable distributions. They proved that:\\

\begin{equation}
\Psi_{\delta}(u)\sim k_{\delta}(1-F(u))
\end{equation}

as $u \to \infty$, \\


where, \\
\indent \indent $u$ = initial surplus\\
\indent \indent $\delta$ = force of interest\\
\indent \indent $k_{\delta}$ = some explicit constant\\

The means that the quotient of the left-hand side and right-hand side tends to $1$ as $u\to\infty$. This could be considered a heavy-tailed counterpart of the work by Sundt and Teugels \cite{sundt1995ruin}. Kalashnikov and Konstantinides \cite{kalashnikov2000ruin} calculated the ruin probability of a risk process with a positive constant force of interest. They used the classical model with the claim process following a Poisson distribution. In the scenario where the claims have different tails, they derived an asymptotic formula for the probability of ruin using advanced arguments.
Epple and Schäfer \cite{epple1996transition} discovered that modeling the claim sizes after an exponential distribution overestimates the ruin probability compared to modeling the claim sizes after the Weibull distribution and the exponential distribution. They suggested a careful approach to modeling claim sizes after an exponential distribution.
However, we will estimate the dependence between claim occurrences considering the required minimum surplus size to prevent the risk of ruin. Even though under the assumptions of dependence, the solutions for the probability of ruin are not straightforward. Valdez and Mo \cite{valdez2002ruin} uncovered that dependence on claims causes ruin to happen much more swiftly.

		\subsubsection{Literature on Dependence between claims}

There is quite a bit of literature on the dependence between insurance claims. Valdez and Mo \cite{valdez2002ruin} provided insight into the statistical time-to-ruin distribution when claims are dependent. They simulated various correlation levels of dependence, specifically at $\eta=0.1$, $\eta=0.2$, $\eta=0.4$, $\eta=0.9$, and $\eta=1$, where $\eta=1$ denotes a strong correlation between the claims. They compared the probabilities of ruin for both dependence and independence with the Lundberg upper bound. They observed that in the event of independence, the Lundberg bound always remains above the probability of ruin, but this is not always true when there is a greater level of dependence. The greater this level of dependence, as measured by the parameter $\eta$ of the copula, the further from this bound it departs. Thus, the existence of correlations between claims is more likely to lead to the ruin of the company than the independence situation, which is often used in practice.
Since the number of claims and the claim sizes are two essential building blocks in the actuarial modeling of insurance claims for non-life insurance, Lee and Shi \cite{lee2019dependent} suggested a dependent modeling structure for joint examination of the two parts of insurance claims (claim number and claim sizes) in a longitudinal context where predictive distribution is the quantity of concern. Using longitudinal dependency modeling, they found that the claim number and claim sizes of these common risks tend to have an elevated serial correlation over time. They gave a structure for modeling recurring insurance claims in a longitudinal configuration using copulas to capture the dependence of the number of claims over time, the dependence of mean claim sizes over time, as well as the dependence between the number of claims and the size of claims. After checking the explanatory variables, they discovered that there is a serial dependence on the frequencies of claims and on the severity over time. By exploiting these dependencies and building a model capturing this dependency, they were able to enhance the forecast of claim scores for frequent kinds of hazards.
Bermudez and Karlis\cite{bermudez2011bayesian} also developed various multivariate Poisson regression models to relax the hypothesis of independence, including zero-inflated models to account for surplus zeros and over-dispersion. To date, these models have been largely ignored, primarily due to their difficulties in computing.
Wu et al.\cite{wu2018aggregate} identified three total claim models with dependence. Model one considers the dependency between indexed insurance benefits induced by a common index; model two takes into consideration the correlation of common fixed costs; model three covers both kinds of dependency. However, they did not test for significance between the various results of both dependence and independence. They only compared them with the Lundberg upper bound probabilities.
But the question is, whether we assume independence or dependence, are the two sets of results significant? This provides a gap to delve into. The significance of the two sets of results can be tested using various statistical tests such as the;
\begin{itemize}
	\item Wilcoxon signed-rank test
	\item Mann–Whitney U test
	\item Friedman test
\end{itemize}

When these statistical tests show a significant difference between the probabilities of independence and the probabilities of dependence, it supports the notion of accepting dependence between the claims. If there is dependence between the claims and a significant difference in the ruin probabilities, then the insurance company can freely assume dependence between the claims and use the dependence models for its analysis. This will give the insurance company a true reflection of its state.

\newpage

\section{METHODOLOGY}
	\subsection{Introduction}
This chapter spells out the methodology that was used to carry out the study. The approach used in this research and other important highlights, such as data and methods of data analysis, are captured in this chapter.
	\subsection{Data}
The data used for this research is secondary data. The data was obtained from insurance companies in Ghana through the National Insurance Commission (NIC). The data includes claims amounts, the number of claims reported, the premium amount, and total premiums received.
	\subsection{Data Analysis}
Data editing and cleaning of the data were done before analyzing the data. This is to ensure accuracy. Diagrammatic presentation by means of tables and graphs was done. Although there were several electronic means of analyzing data, R was used for the analysis. The study adopted copulas to determine the claim dependencies in each insurance product. The probability of ruin, as well as time to ruin, was estimated as a function of the dependence on the number of claims.
	\subsection{Notations and Theorems}
This part consists of Notations and Theorems that were used in this study
		\subsubsection{Notations}
\begin{itemize}
	\item Let $Z_n$ denote the total loss in the unit period $(n-1,n]$. We assume that $\{Z_n, n=1,2,...\}$ is an independent and identically distributed series of random variables with a common distribution function $W(z)$
	\item The Premium is calculated with the loading factor based on the anticipated value principle $\theta > 0$. The constant premium $c=(1+\theta)E(Z_n)$ is paid at the end of each unit period $(n,n-1]$.
	\item 	The insurer surplus at moment n is denoted by $U_n$ and is calculated after paying claims. This same surplus is invested at the beginning of period $(n,n+1]$ at a random interest rate $I_n$.
\end{itemize} 
We assume that\\
\indent  $Eg(Z_1,a)<c(a)$, to avoid the possibility that ruin could occur with probability $1$
		\subsubsection{Theorems}
			\subsubsection*{Theorem 1}
Probability of an insurer’s ruin in finite terms is given recursively as: \\

\begin{equation}
\Psi_1^a(u,i_s)= \sum\limits_{j=1}^{l}P_r\overline{V}[u(1+i_j)+c(a)]
\end{equation}

\begin{equation}
\Psi_{n+1}^a(u,i_s)= \sum\limits_{j=1}^{l}P_sj\{\overline{V}[u(1+i_j)+c(a)]\} + \int\limits_{0}^{u(1+i_s)+c(a)}\Psi_{n}^{a}[u(1+i_j)+c(a)-z,i_j]dV(z)
\end{equation}

\begin{equation}
\Psi_1^a(u,i_s)=P_sj\overline{V}[u(1+i_j)+c(a)]
\end{equation}

Ruin in the $n+1$ periods can only occur in two ways which are mutually exclusive:
\begin{itemize}
	\item[1.] Ruin may occur in the first periods.
	\item[2.] Ruin can occur in one of the periods following the first period if it does not occur in the first period.
\end{itemize}

Since the process $U_n^a$ stationary with independent increment then;\\

\begin{equation}
\Psi_{n+1}^a(u,i_s)= \sum\limits_{j=1}^{l}P_r\int\limits_{0}^{\infty}P_r[\bigcup\limits_{k=1}^{n+1}(U_a^a<0|Z_1^a=z,I_1=i_s)]dV(z)
\end{equation}

\begin{equation}
\Psi_{n+1}^a(u,i_s)= \sum\limits_{j=1}^{l}P_sj\{\overline{V}[u(1+i_j)+c(a)\}+ \int\limits_{0}^{u(1+i_s)+c(a)}\Psi_{n}^{a}[u(1+i_j)+c(a)-z,i_j]dV(z)]
\end{equation}

So, the probability of ruin in an infinite time is obtained by taking a two-sided limit in the above form for $n\to\infty$

\newpage

	\subsection{Surplus Process}
The capital or surplus evolves or changes over time. The classical risk theory is a very relevant stochastic model that helps to understand the process the surplus takes. We can determine the companies reserve at any given time $t$ to be;

\begin{equation}
U(t)=u_0 + \Pi(t)-S(t)
\end{equation}
\indent where,\\
\indent \indent \indent $u_0$ = Initial surplus\\
\indent \indent \indent $\Pi(t)$ = Total Premiums received till time $t$\\
\indent \indent \indent $S(t)$ = Aggregate Claims paid up to time $t$ \\

The surplus process is shown as
\begin{equation}
T = \inf\{t:U(t)<0\}
\end{equation}

If the surplus never becomes zero, $U(t)\geq 0$ for all all $t>0$ then we define $T=\infty$. This is very crucial to define the ruin probabilities.

	\subsection{Aggregate Claims Process}
This study employed two methods in computing the aggregate claim process. The first method is used when the assumption of claim dependency is ignored, and the second is considered when claim dependency is accounted for.
		\subsubsection{Aggregate Claim Process(Independent Assumption)}
Aggregate claims process according to Bowers \cite{bowers1997actuarial}, is given as;
\begin{equation}
S(t)=Z_1+Z_2+...+Z_{N(t)}=\sum\limits_{k=1}^{N(t)}Z_k
\end{equation}
\indent \indent where, \\
\indent \indent \indent \indent $\{S(t):t>1\}$ is the aggregate claim process,\\
\indent \indent \indent \indent $\{N(t):t>0\}$ is the number of claim amount process which is independent of the amount of claims process $Z_k$.

\newpage

		\subsubsection{Aggregate Claim Process (Dependent Assumption)}
Dhaene and Goovaerts \cite{dhaene1997dependency} proposed a model for computing the aggregate claims process with the assumption of claims dependency. The formula is given as;
\begin{equation}
S(t)=\omega_1+\omega_2+...+\omega_n=\sum\limits_{k=1}^{n}\omega_k
\end{equation}
\indent \indent where, \\
\indent \indent \indent \indent $\{S(t):t>1\}$ is the aggregate claim process \\
\indent \indent \indent \indent $n$ is the number of fixed policy holders

	\subsection{Gumbel Copula}
Copulas are functions that connect the marginal distributions to their common multivariate distribution and contain parameters describing the dependence. Gumbel copulas were used in modelling the dependence structure, and it is expressed as;
\begin{equation}
G_\theta(Z,K)=\exp(-[(-\ln{Z})^{\theta}+(-ln{K})^{\theta}]^\frac{1}{\theta})
\end{equation}
\indent \indent where, \\
\indent \indent \indent \indent $Z$ is the distribution claim paid \\
\indent \indent \indent \indent $K$ is the distribution of number of claims $0\leq{Z},K\leq{1}$ and $\theta\in[1,\infty)$.

	\subsection{Ruin Probability}
A useful result for computing ruin probability was adopted and it can be seen in Panjer and Willmot
(1984) as;
\begin{equation}
\psi(\mu)=\frac{\exp^{-R\mu(0)}}{E[\exp^{-R\mu(t)}|T<\infty]}
\end{equation}

$R$ refers to the adjustment coeﬃcient and is the smallest positive solution to
\begin{equation}
E[e^{-r\{S(t)-\Pi(t)\}}]=1
\end{equation}

\newpage

	\subsection{Friedman Test}
Friedman Rank Test is a test statistic to determine whether k groups (ruin probabilities of each product) have been selected from populations having equal medians. 
		\subsubsection{Assumptions}
\begin{itemize}
	\item The data consist of $b$ mutually independent samples (blocks) of $k$ groups. The typical observation $X_{ij}$ is the $jth$  observation in the $ith$  sample (block). 
	\item The interest variable is continuous.
	\item Blocks and treatments do not interact.
	\item The extent of the observation can be classified within each block
\end{itemize}
		\subsubsection{Hypothesis}
$H_0:M_{.1}=M_{.}2=...=M_{.k}$\\
Against the alternative\\
$H_1:$ Not all $M_{.j}$ are equal (where $j=1,2,...,c$)
		\subsubsection{Test Statistic}
To conduct the test, the first step is to replace data values with the corresponding ranks in each of the $k$ independent blocks so that rank 1 is assigned to the lowest value in the block and rank $c$ to the largest. If any values are tied in a block, they are allocated the mean of the ranks they would have been allocated otherwise.\\
\indent The Friedman test statistic is defined as 
\begin{equation}
\chi_r^2=\frac{12}{bk(k+1)}\sum\limits_{j=1}^{k}[R_j-\frac{b(k+1)}{2}]^2
\end{equation}
\indent \indent where, \\
\indent \indent \indent \indent $b$ is mutually independent samples (blocks) \\
\indent \indent \indent \indent $k$ is the number of groups.\\
\indent \indent \indent \indent $R_j$ is the sum of the rank of the $b$ independent blocks in each $k groups$
		\subsubsection{Decision Rule}
Reject $H_0$  at $\alpha$ level of significant if the computed $\chi_r^2$  is greater than or equal to tabulated value $\chi_{1-\alpha}^{2}$ for $k-1$  degrees of freedom.
	\subsection{The Wilcoxon Signed Rank Test}
Wilcoxon Signed Ranks Test is used in comparing two population means in a paired difference experiment. 
		\subsubsection{Assumptions}
\begin{itemize}
	\item The data consist of n values of the difference $D_i$, where each pair of measurement is taken  $(X_i,Y_j)$ 
	\item The difference $D_i$ represent observation on a continuous random variable.
	\item The difference population distribution is symmetrical about their median $M_D$.
	\item The differences are independent. 
	\item Differences are measured at a minimum interval scale. 
\end{itemize}
		\subsubsection{Hypothesis}
$H_0: M_D = 0$\\
$H_1: M_D \neq 0$ \\

For one sided test the hypotheses are presented as\\
$H_0: M_D \geq 0$ for left tail test\\
$H_1: M_D < 0$ \\

and\\
$H_0: M_D \leq 0$ for right tail test\\
$H_1: M_D > 0$
		\subsubsection{Test Statistics}
Let $X$ and $Y$ the population distribution of population 1 and 2 and assumed and we have randomly selected n matched pairs of observations from 1 and 2. Calculate the pairs of difference of the n matched by subtracting each paired population 2 observation from the corresponding population 1 observation and rank the absolute value of the n paired difference from the smallest $(rank1)$ to the largest $(rank n)$. Further, if two or more absolute paired difference are equal, we assign to each ‘tied’ absolute paired difference a rank equal to the average of the consecutive rank that otherwise be assigned to the tied absolute paired difference. \\

\newpage

Let\\
\indent\indent $T^{-}=$ the sum of the rank associated with the negative paired difference\\
and\\
\indent\indent $T^{+}=$ the sum of the rank associated with the positive paired difference\\

The test statistic is given as $T = \min(T^{-},T^{+})$.
		\subsubsection{Decision Rule}
If $H_0$ is true, it implies that the median of the population of difference is 0. For $H_0$ to be rejected the test statistic must satisfies the corresponding rejection point condition: \\
\indent\indent $T\leq T_0$, where\\
\indent\indent\indent\indent $T=\min(T^{-},T^{+})$\\
\indent\indent\indent\indent$T_0$ is the critical value.

\newpage

\section{DATA ANALYSIS AND DISCUSSION}
	\subsection{Introduction}
This section presents the analysis and discussion pertaining to sensitivity analysis of the ruin of insurance companies in Ghana. The analysis includes modelling the number of claims, the amount of claims, estimating the probability of ruin, and the time to ruin with a given surplus value.
	\subsection{Descriptive Analysis}
\textbf{Table 1.1}  represents the mean and standard deviation of the number of claims paid, premiums received, and the amount of claims paid for the study period of an insurance company in Ghana.\\

\textbf{Table 1.1: Descriptive Analysis of Premiums, Claims Paid and Number of Claims} \\

\begin{tabular}{|l|r|r|r|r|}
\hline
 & \textbf{Minimum} & \textbf{Maximum} & \textbf{Mean} & \textbf{Std.Dev} \\ \hline
\textbf{Premiums} & & & & \\ \hline
Motor Insurance	&117,129.59 & 4,087,163.02	&1,592,291.72	&58,918.46 \\ \hline
Householders 	&766.06	& 548,888.75 & 39,647.70 &	17,655.95 \\ \hline
Fire and Allied &	56,017.52	& 251,433.65	& 451,903.53 &	10,4851.20 \\ \hline
Overall	    & 766.06	& 4,087,163.02	& 1,257,523.01	& 655,448.32 \\ \hline
\textbf{Claims Paid} & & & & \\ \hline
Motor insurance &	7,349.00	& 9,940,998.41 &	529,456.00	 & 529,320.19 \\ \hline
Householders &	653.60 &	73,607.80 &	28,671.19 &	28,673.49 \\ \hline
Fire and Allied	& 1,621.00	& 965,520.58 &	242,873.78	& 242,643.31 \\ \hline
Overall	& 653.60	& 9,940,998.41	& 365,932.23 &	367,356.78 \\ \hline
\textbf{Number of Claims} & & & & \\ \hline
Motor insurance &	4.00 &	51.00 &	17.33 &	4.47 \\ \hline
Householders &	1.00 &	17.00 &	8.80 &	2.02 \\ \hline
Fire and Allied &	1.00	&  6.00	& 4.20	& 2.92 \\ \hline
Overall &	1.00& 51.00&	17.53 &	4.68 \\ \hline
\end{tabular}\\

From \textbf{Table 1.1}, it was observed that the mean and the standard deviation of the amount of claim paid per month were 365,932.23 and 365,356.78, respectively. The values range from 653.60 to 9,940,998.41. With respect to policy type, on average, the amount of the claims paid on Motor insurance was 529,456.00 with a standard deviation of 529,320.19, and the values range from 7,349.00 to 9,940,998.41. Also, the mean and the standard deviation of claims paid on Householders insurance product claims were 28,671.19 and 28,673.49, respectively. With respect to Fire and Allied insurance, the mean of the amount of the claims paid was 242,873.78, and the standard deviation was 242,643.31. In general, it can be seen that the mean and the standard deviation are similar.\\
The number of claims reported per month of the insurance company was 17.53 with a standard deviation of 4.68 and ranged from 1 to 51 per month. On average, the number of claims recorded for motor insurance product was 51 with a standard deviation of 4.47. The mean and the standard deviation of claims recorded on Fire and Allied insurance were 4.20 and 2.02, respectively. In general, the mean values are similar to the square of the standard deviation (variance).\\
On average, the insurance company received a premium of 1,257,523.01, ranging from 766.06 to 4,087,163.02. The mean and the standard deviation of premiums received from motor insurance were 1,592,291.72 and 58,918.46, respectively. On average, the company received a premium of an amount of 39,647.70, with a standard deviation of 17,655.95 on Householders insurance product. The Fire and Allied insurance product members contributed on average an amount of 41,903.53 per month.

	\subsection{Modelling Number of Claims}
\textbf{Table 1.2} represents the goodness of fitness test of Poisson distribution on the number of claims recorded per month in the insurance company and policy types. From the descriptive, since the mean and the variance are similar, the possible suggested distribution was Poisson distribution. The chi-square goodness of fit was then performed to determine if the number of claims follows a Poisson distribution and the results were presented in Table 1.2.\\

\textbf{Table 1.2: Model Adequacy for Number of Claims}\\
\begin{tabular}{|l|c|c|c|}
\hline
\textbf{Policies} & \textbf{Chi-Square} & \textbf{Degree of Freedom} & \textbf{p-value}\\ \hline
Motor Insurance & 48.845 & 58 & 0.7986 \\ \hline
Householders & 52.452 & 58 & 0.6808 \\ \hline
Fire and Allied & 49.456 & 58 & 0.7803 \\ \hline
Overall & 45.652 & 58 & 0.8803 \\ \hline
\end{tabular} \\

The rate of the claims paid per month was estimated using a Maximum Likelihood Estimator (MLE), and the results can be seen in \textbf{Table 1.3}\\

\pagebreak
\textbf{Table 1.3: Modelling Amount of Claims}\\
\begin{tabular}{|l|c|c|}
\hline
\textbf{Policies} & \textbf{Rate($\hat{\vartheta}$)} & \textbf{Standard Error} \\ \hline
Motor insurance &	0.000000573	& 0.000006879 \\ \hline
Householders 	& 0.000000320	& 0.000000024 \\ \hline
Fire and Allied	& 0.000000094	& 0.000000045 \\ \hline
Overall	& 0.000002730	& 0.000000359 \\ \hline

\end{tabular} \\

From Table 1.3, it can be seen that the amount of claims paid is at the rate of 0.00000273 per month. With respect to the policies, Motor Insurance recorded the highest rate of claims paid ($\hat{\vartheta}=0.000000573$), followed by Fire and Allied ($\hat{\vartheta}=0.00000032$).

	\subsection{Estimating Dependency Between Claims Amount and Number of Claims}
One of the objectives of the study was to measure the dependencies that exist between the number of claims recorded and the amount of claims paid. This test was done using Pearson Correlation ($r$) and the Copula test of dependency ($\eta$).\\

\textbf{Table 1.4: Estimating dependency between Claims Amount and Number of Claims}\\

\begin{tabular}{|l|c|c|c|c|}
\hline
\textbf{Company} & \textbf{Pearson Correlation $r$} & \textbf{P-value} & \textbf{Test of dependence $\eta$} & \textbf{P-value} \\ \hline
Motor insurance	& 0.827	& 0.000	& 0.074	& 0.8520	\\ \hline
Householders 	& 0.858	& 0.000	& 0.046	& 0.8245	\\ \hline
Fire and Allied	& 0.891	& 0.000	& 0.031	& 0.8003	\\ \hline
Overall	& 0.757	& 0.000	& 0.147	& 0.7265 \\ \hline
\end{tabular} \\

A low value of the copula implies a higher dependency or correlation between variables. From Table 1.4, with regards to the correlation coefficient, it can be seen that all the p-values are less than 0.05. Hence, there exists a significant positive relationship between the amount of claims paid and the number of claims recorded. With regards to the copula test of dependency, since the p-values were greater than 0.05, the null hypothesis that there is a claim dependency was not rejected. Among the various products, Fire and Allied insurance recorded the highest level of dependency with a correlation coefficient ($r$ = 0.891; $p-value$ = 0.031) and a lower copula value ($\eta$ = 0.031; $p-value$ = 0.8003).

	\subsection{Probability of Ruin}
Sensitivity analysis of the start-up capital (initial surplus) of the insurance company to the risk of ruin was done in the case of claim independence and dependence. The results are presented in Table 1.5.\\	
	
\textbf{Table 1.5: Probabilities of Ruin} \\	

\begin{tabular}{|l|l|c|c|}
\hline
\textbf{Initial Surplus} & \textbf{Product} & \textbf{Dependence} & \textbf{Independence} \\ \hline
\textbf{0} & Motor insurance &	0.318800 &	0.310800 \\ \hline
 & Householders &	0.342200 &	0.333500 \\ \hline
 & Fire and Allied	& 0.346300	& 0.338500 \\ \hline
 & Overall &	0.317970 &	0.309600 \\ \hline
\textbf{500} & Motor insurance &	0.255040 &	0.244400 \\ \hline
 & Householders & 	0.273760 &	0.262450 \\ \hline
 & Fire and Allied &	0.277040 &	0.263432 \\ \hline
 & Overall	& 0.239760 &	0.226630 \\ \hline
\textbf{1000} & Motor insurance &	0.223160 &	0.210360 \\ \hline
 & Householders & 	0.239540 &	0.225446 \\ \hline
 & Fire and Allied &	0.242410 &	0.223838 \\ \hline
 & Overall &	0.197790 &	0.182418 \\ \hline
\textbf{1500} & Motor insurance &	0.207220 &	0.176820 \\ \hline
 & Householders &	0.222430	& 0.189283 \\ \hline
 & Fire and Allied	& 0.225095 &	0.184753 \\ \hline
 & Overall &	0.169805 &	0.141425 \\ \hline
\textbf{2000} & Motor insurance &	0.165776 &	0.125344 \\ \hline
 & Householders &	0.177944 &	0.134853 \\ \hline
 & Fire and Allied &	0.180076 &	0.130811 \\ \hline
 & Overall	& 0.127844 &	0.077950 \\ \hline
\textbf{2500} & Motor insurance &	0.132621 &	0.083981 \\ \hline
 & Householders &	0.142355 &	0.088657 \\ \hline
 & Fire and Allied &	0.144061 &	0.084266 \\ \hline
 & Overall &	0.104275 &	0.060262 \\ \hline
\textbf{3000} & Motor insurance &	0.106097 &	0.041405 \\ \hline
 & Householders &	0.113884 &	0.044077 \\ \hline
 & Fire and Allied &	0.115249 &	0.044447 \\ \hline
 & Overall &	0.100902 &	0.041202 \\ \hline
\textbf{3500} & Motor insurance &	0.084877 &	0.004053 \\ \hline
 & Householders &	0.091107 &	0.005116 \\ \hline
 & Fire and Allied &	0.092199 &	0.006860 \\ \hline
 & Overall &	0.084160 &	0.002399  \\ \hline

\end{tabular}\\

\pagebreak

From Table 1.5, it can be seen that at any initial surplus, the probability of ruin in the case of claim dependence was higher than that in the case of claims independence. Also, with higher initial surplus values, the lower the probability of ruin and vice versa. For higher levels of surplus, this leads to a lower probability of ruin. Apparently, the initial surplus provides some capacity to absorb shocks from expectations, particularly in claims. This observation is true across various insurance products. For example, when the surplus is 500, the probability of ruin in the case of claim dependence for motor and householders’ insurance was 0.255040 and 0.273760, respectively. At an initial surplus of 2000, the probability of ruin in the case of claim dependence for motor and householders’ insurance reduces to 0.165776 and 0.125344, respectively.

	\subsection{Testing for the Difference Between the Ruin Probability on Claims Dependent and Claims Independent}
From Table 1.4, it was found that the data exhibit a significant level of claim dependency. As a result of that, a test was performed to determine the effect of assuming claim independence in estimating the ruin probability when in fact, they are claim dependent. A Wilcoxon signed rank test was done, and the results were presented in Table 1.6. \\

\textbf{Table 1.6: Wilcoxon Signed Rank test on Ruin Probabilities} \\

\begin{tabular}{|l|c|c|c|c|c|c|}
\hline
\textbf{Product} & \textbf{Neg. Rank} & \textbf{Mean Rank} & \textbf{Sum of Ranks} & \textbf{Test Statistic} & \textbf{P-value} \\ \hline
Motor insurance & 8 &	4.50 &	36.00 &	-2.521 &	.012 \\ \hline
Householders	& 8	& 4.50 &	36.00 &	-2.521	& .012 \\ \hline
Fire and Allied &	8 &	4.50 &	36.00 &	-2.521 &	.012 \\ \hline
Overall &	8 &	4.50 &	36.00 &	-2.521 &	.012 \\ \hline

\end{tabular} \\

From Table 1.6, we see that in all the products and the company in general, there was a significant difference between the ruin probability for claims dependent and claims independent. This implies that assuming claim independence without testing for claim dependence in computing ruin probabilities may lead to misleading results.

	\subsection{Plotting Ruin Probability Based on Dependent and Independent Claims Assumptions Against Start-up Capital}
	
A plot of the probability of ruin in the case of claim dependence and claims independence was created against initial capital. This was done to establish the relationship between start-up capital and the level of dependence. The plot can be seen in Figure 1.1, 1.2, 1.3, and 1.4 for motor, Householders, Fire and Allied, and the company in general.
All the figures exhibit similar behaviors. As the initial surplus increases in all the graphs, the ruin probabilities for both claim dependence and independence decrease. As the initial surplus increases, ruin probabilities in the case of independence are much higher than those indicated by the case of dependence. Hence, the gaps between ruin probabilities in the case of independence and dependence become wider. Since Fire and Allied insurance exhibits the highest dependencies among the various insurance products, the Fire and Allied insurance displays the widest gaps, and it can be seen in Figure 1.3. \\

\includegraphics[scale=1]{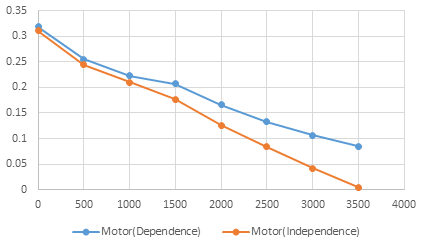} \\
\textbf{Figure 1.1: Motor Insurance ruin probabilities against capital}\\

\includegraphics[scale=1]{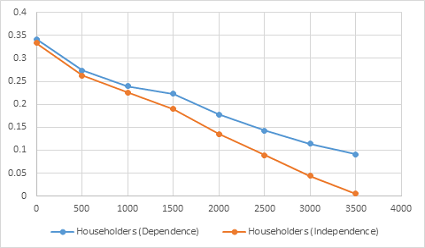} \\
\textbf{Figure 1.2: Householders ruin probabilities against capital} \\

\pagebreak

\includegraphics[scale=1]{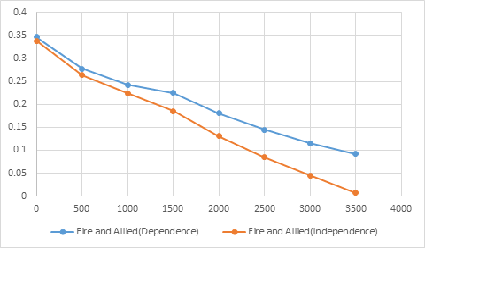} \\
\textbf{Figure 1.3: Fire and Allied ruin probabilities against capital} \\

\includegraphics[scale=1]{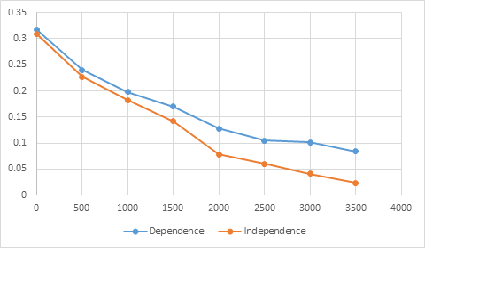}\\
\textbf{Figure 1.4: Overall Company's ruin probabilities against capital} \\

	\subsection{Analysis of Product Performance}
From Table 1.5, it was established that there exists claims dependency--meaning that assuming claim independence is not correct. Hence, to determine the best product, a Friedman test was performed on the ruin probabilities of the various products using the assumption of claim dependencies. The results can be seen in Table 1.7. \\

\pagebreak

\textbf{Table 1.7: Friedman Test on the three Insurance Products}\\
\begin{tabular}{|l|c|c|c|c|}
\hline
\textbf{Product} & \textbf{Mean Rank} & \textbf{Median} & \textbf{Test Statistic} & \textbf{P-value} \\ \hline
Motor Insurance & 1 & 0.186498 & 16	& 0.000 \\ \hline
Householders	& 2	& 0.200187 & 16	& 0.000 \\ \hline
Fire and Allied	& 3	& 0.202586 & 16	& 0.000 \\ \hline 
\end{tabular} \\

From Table 1.7, it can be seen that the median of the ruin probabilities for Motor Insurance (Med=0.1865), Householders Insurance (Med=0.2002), and Fire and Allied Insurance (Med=0.2026) were significantly different. It was observed that among all the products, Motor Insurance had the lowest ruin probability. This implies that if the company were to offer only Motor Insurance, the company would survive longer than offering any other insurance product. Therefore, among the three products, Motor Insurance is the best performing product in the insurance industry, followed by Householders Insurance and then Fire and Allied Insurance. The difference between the ruin probabilities of the insurance products was tested using the Wilcoxon matched-paired test, and the results were presented in Table 1.8.\\

\textbf{Table 1.8: Differences in Ruin Probability per Insurance Product}\\
\begin{tabular}{|l|c|c|c|c|c|c|}
\hline
\textbf{Products} & \textbf{Rank} & \textbf{No.} & \textbf{Mean Rank} & \textbf{$\Sigma$ Ranks} & \textbf{Test.stat} & \textbf{P-value} \\ \hline
HouseH v. Motor &	Negative Ranks	& 0 &	0.00 &	0.00 &	-2.521 &	0.012 \\ \hline
 & Positive Ranks &	8 &	4.50 &	36.00 &	& \\ \hline	
 & Total &	8 & & & & \\ \hline
FireA v. Motor &	Negative Ranks	& 0 &	0.00 &	0.00 &	-2.521 &	0.012 \\ \hline
 & Positive Ranks &	8 &	4.50 &	36.00 &	& \\ \hline	
 & Total &	8 & & & & \\ \hline
FireA v. HouseH &	Negative Ranks	& 0 &	0.00 &	0.00 &	-2.521 &	0.012 \\ \hline
 & Positive Ranks &	8 &	4.50 &	36.00 &	& \\ \hline	
 & Total &	8 & & & & \\ \hline
\end{tabular} \\

From Table 1.8, comparing the Householders Insurance and Motor Insurance products, it was observed that there was a significant difference between them with 8 positive ranks. This implies that at the same initial surplus, the ruin probabilities of Householders Insurance products are greater than those of Motor Insurance products. With regard to Fire and Motor Insurance, it was also found that the ruin probabilities of Fire Insurance policies were significantly greater than the Motor Insurance policies. Focusing on Fire and Householders Insurance policies, the study revealed that the probabilities of ruin recorded for Fire policies are significantly higher than those of Householders Insurance.

\newpage

\section{SUMMARY, CONCLUSION AND RECOMMENDATIONS}
	\subsection{Summary}
Insurance companies are risk bearers; however, they are also exposed to the likelihood of running into ruin. This is the situation where the initial surplus falls below zero. There is a need to find the required start-up capital to hedge against insolvency. This study determines the sensitivity of the start-up capital (initial surplus) of the insurance company on the probability of ruin and also determines the dependence in the claims of various insurance policies. The study reviews literature concepts of ruin theory and start-up capital. Secondary data from an insurance company was extracted through the National Insurance Commission (NIC) for the period of 2013 to 2017. The data includes claims amounts, the number of claims reported, premium amounts, and total premiums.

	\subsection{Summary of Findings}
The study revealed that the number of claims recorded in the insurance company follows a Poisson distribution with a rate of 17.7167. Moreover, with respect to various insurance policies, claims counts for motor insurance, fire and allied insurance, and householders insurance follow a Poisson distribution with rates of 18.4833, 8.8333, and 4.4500 respectively. With respect to the amounts of claims paid, the study found that the amount of claims paid follows an exponential distribution with a rate of 0.00000273 per month. Motor Insurance recorded the highest rate of claims paid of 0.000000573 per month.\\
The study also reveals that there exist claims dependencies in the insurance company and among the various products, Fire and Allied insurance recorded the highest level of dependency with a correlation coefficient ($r$=0.891, p-value=0.031) and a lower copula value ($\eta$=0.031, p-value=0.8003).\\
The study found that when there is dependency of claims, the probabilities of ruin computed based on the assumption of independence of claims are significantly different from the ruin probabilities computed on the assumption of claim dependency. It was also found that among the three insurance policies, the probability of ruin for motor insurance policies was the lowest. This means that motor insurance policies are performing well in the insurance companies.\\
The study also revealed that when the initial surplus is high, the ruin probability decreases. When the initial surplus increases, the ruin probabilities in the case of independence are much higher than those indicated by the case of dependence. This leads to a wider difference between them with increasing surplus start-up.

	\subsection{Conclusion}
The following conclusions were drawn from the study: There exists a dependency between the number of claims and the amounts of claims in the insurance company’s claims for the various insurance products. Fire and Allied Insurance exhibit the highest dependence among the general insurance products in Ghana. Analyzing the performance of the various insurance policies, the study concluded that motor insurance policies are performing very well. All these findings imply that when there is dependence in the claim data, computing the ruin probability based on the assumption of independence results in underestimation. According to the sensitivity analysis, the gap of underestimation widens as the start-up capital increases.

	\subsection{Recommendations}
\begin{itemize}
	\item Since the gap of underestimation between the ruin probabilities (when claims are dependent and claims are assumed to be independent) widens as the initial capital increases, large companies with large reserves should consider assuming dependence between the claims data to avoid having misleading results. 
	\item The Fire and Allied insurance policy recorded the highest ruin probability. The insurance companies with more Fire and Allied insurance policyholders should raise more awareness of the perils and consequences of fire outbreaks to the general public to reduce the risk of such perils occurring frequently. Thus, the rate at which claims are reported will be reduced to obtain a safer probability of ruin.
\end{itemize}

\pagebreak

\section*{Appendix}
	\subsection*{R codes}

library(MASS) \\
library(gumbel)\\
library(copula)\\

\noindent dannyp=read.delim("clipboard") \# loading the data \# \\
names(dannyp) \# Variables names \# \\
motorc=dexp(dannyp \$motor, rate = 1, log = FALSE)\\

\noindent \# Poisson\# \\
motora<-fitdistr(dannyp\$motora, "Poisson") \\
firea<-fitdistr(dannyp\$firea, "Poisson") \\
assetda<-fitdistr(dannyp\$accda, "Poisson") \\
overall<-fitdistr(dannyp\$overall, "Poisson") \\

\noindent \# Estimators of elliptical copula parameters \# \\
library(qrmtools)\\
X = returns(SMI.12) \# compute log-returns\\
U = pobs(X) \# compute pseudo-observations\\
d = ncol(U) \# 20 dimensions\\

\noindent f.irho = fitCopula(normalCopula(dim = d, dispstr = "un"), data = U, method = "irho") \\
f.itau = fitCopula(normalCopula(dim = d, dispstr = "un"), data = U, method = "itau")\\

\noindent P.irho = p2P(coef(f.irho), d = d)\\
P.itau = p2P(coef(f.itau), d = d)\\

\noindent plot(density(x), col="goldenrod", lwd=1, bty="n", \\
\indent    xlim=c(0,20), ylim=c(0,0.35), \\
\indent    xlab=expression($chi^{\{squared\}}$), \\
\indent    ylab="Probability", \\
\indent    main="The Chi-square distribution with df=(3, 4, 5, and 6)") \\
abline(h=0.05, col="red", lty=6) \\
abline(h=0.01, col="red", lty=3)\\
par(new=TRUE)\\
lines(density(x1), col="pink", lty=2)\\
par(new=TRUE)\\
lines(density(x2), col="brown", lty=1)\\
par(new=TRUE)\\
lines(density(x3), col="dodgerblue4", lty=5)\\

\noindent observed = table(shad\$numberOfParasites); observed\\
mean(shad\$numberOfParasites)\\
sum(shad\$numberOfParasites)\\

\noindent x = seq(0, 6, by=1); x \\
x.d = dpois(x, lambda=0.9453782); x.d\\
sum(x.d)\\
prob.1 = x.d[1:6]; prob.1\\
prob.2 = 1 - sum(prob.1); prob.2\\
probs = c(prob.1, prob.2); probs\\
sum(probs)\\
expected = round(probs*238, digits=2); expected \\
cbind(observed, expected)\\

\noindent new.observed = c(103, 72, 44, 14, 5)\\
new.probs = c(probs[1:4], sum(probs[5:7]))\\
Xsq = chisq.test(new.observed, p=new.probs); Xsq\\

\noindent pchisq(57.245, df=58, lower.tail=FALSE)\\
pchisq(55.425, df=58, lower.tail=FALSE)\\
pchisq(60.452, df=58, lower.tail=FALSE)\\
pchisq(65.567, df=58, lower.tail=FALSE)\\

\noindent ex1=rexp(60, rate=529456.00)\\
fitdistr(ex1, "exponential")\\
mean(ex1)\\

\noindent pchisq(48.845, df=58, lower.tail=FALSE)\\
pchisq(52.452, df=58, lower.tail=FALSE)\\
pchisq(49.456, df=58, lower.tail=FALSE)\\
pchisq(45.652, df=58, lower.tail=FALSE)\\

\noindent mt1=rpois(60, lambda=17.33)\\
fitdistr(mt1, "Poisson")\\

\noindent mt2=rpois(60, lambda=8.80)\\
fitdistr(mt2, "Poisson")\\

\noindent mt3=rpois(60, lambda=4.20)\\
fitdistr(mt3, "Poisson")\\

\noindent mt4=rpois(60, lambda=17.53)\\
fitdistr(mt4, "Poisson")

\vspace{1cm}
\newpage

\bibliographystyle{plain}
\bibliography{mybib}

\end{document}